\documentclass[reprint, groupedaddress, amsmath, amssymb, aps]{revtex4-2}
\usepackage{graphicx}
\usepackage{dcolumn}
\usepackage{bm}
\usepackage{subfigure}
\usepackage{appendix}

\begin{document}

\title{On Density Limit of Lower Hybrid Current Drive caused by Parametric Decay Instability in Tokamak Plasmas}

\author{Kunyu Chen}
\affiliation{Department of Engineering Physics, Tsinghua University, Beijing 100084, China}

\author{Zhihao Su}
\affiliation{Department of Engineering Physics, Tsinghua University, Beijing 100084, China}

\author{Zikai Huang}
\affiliation{Department of Engineering Physics, Tsinghua University, Beijing 100084, China}

\author{Long Zeng}
\affiliation{Department of Engineering Physics, Tsinghua University, Beijing 100084, China}

\author{Zhe Gao}
\email{gaozhe@tsinghua.edu.cn}
\affiliation{Department of Engineering Physics, Tsinghua University, Beijing 100084, China}

\date{\today}

\begin{abstract}
The density limit of lower hybrid current drive (LHCD) is scaled by coupling the saturation process of parametric instability induced by LH waves in the scrape off layer (SOL) plasma to the propagation of waves. It is shown that the density limit of LHCD satisfies $n_\text{lim}\propto L_y^{2/3} P_0^{-2/3}\omega_0^{2}B_0^{2/3}T_e$, which is consistent with results of simulations and previous experiments. Both theoretical analysis and simulation results indicate that the density limit is far from being reached for ITER baseline profile. Therefore, the density limit phenomena will not prevent LHCD from being a promising method of driving plasma current at ITER and future tokamaks. 
\end{abstract}

\maketitle


Effective non-inductive current drive in high-density plasma is essential for sustainable fusion energy production on tokamaks. Among all available methods, lower hybrid current drive (LHCD)\cite{Fisch_1980,Fisch_1987} has the highest efficiency of driving plasma current as well as controlling the current profile and improving energy confinement\cite{Hawkes_2001,EAST_SciAdvances,Fujita_2000}. 
However, LHCD faces serious challenge of density limit, an anomalous efficiency loss at a plasma density much lower than the classical limit of wave accessibility, which has been attributed to wave-plasma interactions in the scrape-off layer (SOL) plasma. The density limits of LHCD observed on various tokamaks including JET\cite{Cesario_2006,Cesario_2011}, EAST\cite{Ding_2017,Li_2022}, Alcator C-Mod\cite{Wallace_2010,Wallace_2011}, FTU\cite{Ridolfini_2011} and Tore Supra\cite{Goniche_2013} are significantly lower than the required level for ITER, which makes the effectiveness of LHCD questionable on ITER and future fusion reactors. 

Various efforts have been made to understand and overcome the density limit of LHCD. On FTU\cite{Cesario_2010}, efficient LHCD at a plasma density comparable to the ITER level is achieved by using a lithium-coated vessel to increase the electron temperature of SOL plasma, which results in suppression of parametric decay instability (PDI)\cite{Zhao_2013_1,Cesario_2006}. On Alcator C-Mod, PDI can be suppressed by increasing the toroidal plasma current to adjust the SOL plasma density profile\cite{Baek_2018}, thus leads to a better LHCD efficiency at high plasma density. Recent experiments on EAST\cite{Li_2016,Li_2022} using LH antennas at different frequencies show that the density limit of LHCD can be improved by increasing LH frequency and toroidal magnetic field. However, despite all the experimental efforts to improve LHCD efficiency at high density plasma, there remains an absence of a comprehensive theoretical framework capable of unifying the experimental findings. For the most likely candidate, PDI process, previous studies about the linear growing stage\cite{Porkolab_1974,Liu_1986,Cesario_2010} and the convective saturation\cite{Chen_1977,Chen_1977_1,Liu_1984,Cesario_2006,Takase_1983} mainly focus on analysis of the PDI decay channel. Therefore, when it comes to problem of density limit, only qualitative explanations are provided. Consequently, we are still unable to provide an effective prediction of the availability of LHCD on ITER and future fusion reactors. In addition to the problem of unexpected spectral upshift\cite{Peysson_2016}, LHCD is now absent from the heating and current drive plan of ITER, which makes its steady-state operation much more challenging.

This letter shows that the gap of understanding the LHCD density limit can be filled by coupling the PDI process arises from plasma fluctuation to the propagation of LH waves. A quantitative theoretical analysis on the power loss of the injected LH wave in SOL plasma as well as a simulation of self-consistent evolution of the spectrum and power flux of the LH wave are presented. The scaling law of LHCD density limit is acquired, and the predicted convective loss in the SOL plasma is minimal for an ITER baseline profile. 

\begin{figure}[h]
    \centering
    \includegraphics[width=1\linewidth]{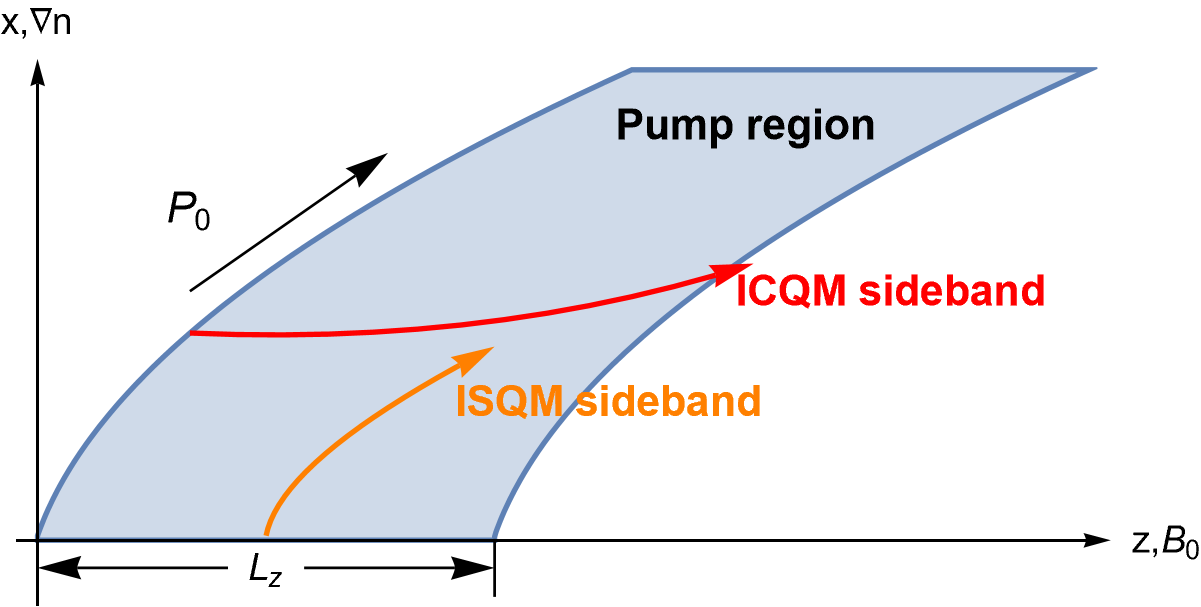}
    \caption{Pump region and typical sideband trajectories}
    \label{fig1}
\end{figure}

For a typical LH induced PDI process, the LH wave launched from the antenna serves as pump wave and decays into a lower sideband LH wave and a low-frequency quasi-mode. The coupling equations in an inhomogeneous plasma are\cite{Rosenbluth_1972}
\begin{equation}
  \begin{aligned}
    \left[\epsilon_1
    +\frac{\partial \epsilon_1}{\partial \omega_1}\cdot i\frac{\partial}{\partial t}
    -\frac{\partial \epsilon_1}{\partial k_{1x}}\cdot i\frac{\partial}{\partial x}\right]\phi_1
    =&\alpha_{L\rightarrow 1}\phi_0^*\phi_L\\
    \left[\epsilon_L
    +\frac{\partial \epsilon_L}{\partial \omega_L}\cdot i\frac{\partial}{\partial t}
    -\frac{\partial \epsilon_L}{\partial k_{Lx}}\cdot i\frac{\partial}{\partial x}\right]\phi_L
    =&\alpha_{1\rightarrow L}\phi_0\phi_1\\
  \end{aligned}
  \label{nonlocal1}
\end{equation}
where $j=0,1,L$ refers to the pump, the lower sideband and the low-frequency quasi-mode, $\epsilon_j$ refers to the dielectric function of wave $j$, $\alpha_{i\rightarrow j}$ represents the coupling coefficient from daughter wave $i$ to daughter wave $j$. For typical SOL plasma, the dominant saturation mechanism of PDI is the finite profile of the pump wave\cite{Chen_2024_1}, and the effects of wavenumber mismatch due to plasma inhomogeneity is minimal. Thus, the amplification factor of PDI takes the form of

\begin{equation}
    A_\text{amp}=\int \gamma_0 dt
    \label{intA}
\end{equation}
Such integration is proceeded along the trajectory of the sideband LH wave, where $dt=\frac{dl}{v_{1g}}$ with $v_{1g}$ the group velocity and $\gamma_0$ is the linear growth rate of PDI. FIG.~\ref{fig1} shows the typical pump region and sideband trajectory for LHCD, in which $z$ is the toroidal direction with $L_z$ the toroidal scale of the LH antenna and $x$ the radial direction. 

Integration Eq.~\eqref{intA} can be simplified by introducing the conservation of pump power flux $P_0$. We choose a sideband wave propagates across the pump region to maximize $A_\text{amp}$, thus the pump power flux across the sideband trajectory equals to total power flux $P_0$ when pump depletion effect is neglected, namely 
\begin{equation}
    \begin{aligned}
        P_0&=\int W_0 L_y (v_{0gx}dz-v_{0gz}dx)\\
        &=\int W_0 L_y \left(v_{0gx}v_{1gz}-v_{0gz}v_{1gx}\right)dt
    \end{aligned}
    \label{P0}
\end{equation}
where $W_0$ is the power density of pump and $L_y$ the poloidal scale of the LH antenna. Substituting Eq.\eqref{P0} to simplify the integration in Eq.\eqref{intA} and noticing that $P_0=W_0 L_y L_z v_{0gx}$ is always satisfied, we find
\begin{equation}
    A_\text{amp}=\left\langle\frac{\gamma_0 L_z}{v_{1gz}-v_{0gz}v_{1gx}/v_{0gx}}\right\rangle
    \label{Aavg}
\end{equation}
the brackets in Eq.\eqref{Aavg} refer to the average value along the sideband trajectory. 

There are two main decay channels for LHCD induced PDI in SOL plasma. The decay channel involving an ion-sound quasi-mode (ISQM) dominates PDI process near the antenna mouth with low plasma density $n_e<10^{18}/\text{m}^3$\cite{Cesario_2006}, whereas the channel involving an ion-cyclotron quasi-mode (ICQM) gets dominant near the last closed flux surface (LCFS) with plasma density $n_e\sim 10^{19}/\text{m}^3$\cite{Zhao_2013_1}. Both channels can achieve notable linear growth rate $\gamma_0 >\omega_{ci}$, but their effects on density limit is totally different. The ISQM channel has no direct contribution to density limit since its linear growth rate usually decreases with plasma density\cite{Cesario_2006}, whereas the growth rate of an ICQM channel increases with plasma density\cite{Zhao_2013_1}. Therefore, we consider only the convective loss caused by ICQM channels to evaluate the density limit of LHCD. Note that the ICQM channels reach the largest growth rate at scatter angles $\delta=\angle \mathbf{k}_{0\perp}\mathbf{k}_{1\perp}\approx\frac{\pi}{2}$, here footnote $\perp$ refers to the direction perpendicular to the magnetic field. Since the pump LH wave has $k_{0y}\approx0$, we then find the sideband wave with the largest $A_\text{amp}$ has $k_{1x}\approx 0$. Consider the group velocity and dispersion relation of LH wave
\begin{equation}
    \begin{aligned}
         v_{jg\perp}&=\frac{m_i}{m_e}\cdot\frac{\omega_{LH}^2}{\omega_j}\cdot\frac{k_{jz}^2 k_{j\perp}}{k_j^4}\\
         \frac{v_{jgz}}{v_{jg\perp}}&=\frac{k_{j\perp}}{k_{jz}}=\sqrt{\frac{m_i}{m_e}\cdot\frac{\omega_{LH}^2}{\omega_{j}^2-\omega_{LH}^2}}
    \end{aligned}    
    \label{vgLH}
\end{equation}
where $j=0,1$ refers to the pump and sideband LH waves and $\omega_{LH}=\frac{\omega_{pi}^2}{1+\omega_{pe}^2/\omega_{ce}^2}$ the lower hybrid resonant frequency. According to Eq.\eqref{vgLH}, ICQM channels always satisfy $v_{1gz}\gg v_{1g\perp}$, thus we may neglect the radial scale of the sideband trajectory of an ICQM channel and treat $\gamma_0$ as a constant. Applying $v_{jgx}\ll v_{jgz}$ and $k_{1x}\approx 0$ to Eq.\eqref{Aavg} and considering $v_{1gz}=\frac{c}{n_{1z}}$ where $c$ is the light speed and $n_{1z}$ the refraction factor of the sideband along the magnetic field, we find
\begin{equation}
    A_\text{amp}=n_{1z}\gamma_0\cdot\frac{L_z}{c}
    \label{A0}
\end{equation}
can be applied to evaluate the convective loss, which takes the form of
\begin{equation}
    P_\text{loss}=W^{(0)}e^{2 A_\text{amp}}\cdot v_{1gz} L_y L_x
\end{equation}
where $W^{(0)}$ is the power density of the low-frequency fluctuation serves as the initial state of PDI, and $L_x$ refers to the radial scale of the SOL region. For simplicity, we define the density limit of LHCD as the critical density at which  $P_\text{loss}=0.8 P_0$, i.e. 80\% of the pump power flux converts to the sideband LH wave through ICQM decay in the SOL region. Previous literature has stated that the power density of plasma fluctuation $W^{(0)}$ is about 8 magnitudes lower than the pump wave power density $W_0$\cite{Cesario_2006,Chen_1977_1}, thus the critical amplification factor is approximately $A_\text{amp}^\text{cr}=12$. The differences in the linear terms $v_{1gz}$ and $L_y, L_x$ have minimal effect to the critical amplification factor. Since the amplification factor is a function of plasma density, the density limit $n_\text{lim}$ can be defined by $A_\text{amp}\left(n_\text{lim}\right)=12$.

Fortunately, the relationship between the amplification factor $A_\text{amp}$ and the plasma and LH parameters is rather simple at typical SOL parameter near LCFS. Investigations of such relationship can be completed analytically with limited numerical assistance. 
The linear growth rate $\gamma_0$ satisfies
\begin{equation}
    \gamma_0\approx \text{Re}\left[\frac{-i\alpha_{1\rightarrow L}\alpha_{L\rightarrow 1}\left|\phi_0^2\right|}{\epsilon_L \left(\partial\epsilon_1/\partial\omega_1\right)}-\nu_1\right]
    \label{gamma0}
\end{equation}
with $\nu_1=\frac{i\epsilon_1}{\partial \epsilon_1/\partial \omega_1}$ the linear damping rate of the sideband. Consider the power density of the pump wave $W_0=\frac{1}{2}\epsilon_0 k_0^2 |\phi_0|^2$, thus we find $\left|\phi_0\right|^2\propto \frac{1}{v_{0gx}}$ since $P_0$ and $L_y, L_z$ are fixed. The kinetic coupling coefficients of LHCD induced PDI considering both $E_\parallel$ and $E\times B$ coupling are\cite{Liu_1986,Zhao_2013_1}
\begin{equation}
    \begin{aligned}
         \alpha_{1\rightarrow L}&=\frac{e\phi_0}{4m_e}\cdot\frac{\chi_{eL}}{k_{Lz}}\\
         &\left[-\frac{ik_{0\perp}k_{1\perp}}{\omega_{ce}}\left(\frac{k_{0z}}{\omega_1}-\frac{k_{1z}}{\omega_0}\right)+k_{0z}k_{1z}\left(\frac{k_{0z}}{\omega_1^2}+\frac{k_{1z}}{\omega_0^2}\right)\right]\\
         \alpha_{L\rightarrow 1}&=\frac{e\phi_0}{2m_e}\cdot\frac{k_L^2}{\omega_1 k_{1}^2}\cdot\left(1+\chi_{iL}\right)\left[-\frac{ik_{0\perp}k_{1\perp}}{\omega_{ce}}+\frac{k_{1z}k_{0z}}{\omega_0}\right]        
    \end{aligned}
    \label{couplingcoefficients}
\end{equation}
here $\angle \mathbf{k}_{1\perp} \mathbf{k}_{0\perp}=\frac{\pi}{2}$ is assumed for such decay channels have the largest growth rate\cite{Zhao_2013_1}, $\chi_{iL,eL}$ are the susceptibilities for ICQM with the following form
\begin{equation}
    \chi_{sL}=\frac{\omega_{ps}^2}{k_L^2 v_s^2}\left[1+\zeta_{s}^{(0)}\sum_{n=-\infty}^{+\infty}Z\left(\zeta_{s}^{(n)}\right)I_n\left(b_s\right)e^{-b_s}\right]
    \label{chi}
\end{equation}
Here $\zeta_{s}^{(n)}=\frac{\omega_L-n\omega_{cs}}{k_{Lz}v_s}$, $\omega_{cs}, \omega_{ps}$ and $v_s$ refers to the gyro-frequency, electrostatic oscillation frequency and thermal velocity of component $s$, $b_{s}=k_{L\perp}^2\rho_s^2$ satisfies $b_{e}\ll 1$ and $b_i\sim 1$. Previous studies\cite{Liu_1984,Liu_1986,Cesario_2006,Zhao_2013_1} have shown that $E\times B$ coupling is the main coupling mechanism for the PDI scenarios with high plasma density $n_e>10^{19}/\text{m}^3$, thus we consider only $E\times B$ coupling in the following derivation. Substituting Eq.\eqref{vgLH} to Eq.\eqref{gamma0}, Eq.\eqref{couplingcoefficients} and Eq.\eqref{chi}, assuming $T_i=T_e$ and $|\omega_0|\approx|\omega_1|\gg|\omega_L|, n_{1z}\approx n_{Lz}\gg n_{0z}$, then Eq.\eqref{A0} is reduced to the following form 
\begin{equation}
    \begin{aligned}
        A_\text{amp}\propto \omega_{LH}^3 P_0 L_y^{-1} T_e^{-1} \omega_0^{-2} B_0^{-2}
        \cdot n_{1z} F\left(\zeta_s^{(n)},b_i\right)
    \end{aligned}    
    \label{scaleF}
\end{equation}
where function $F$ takes the form of
\begin{equation}
    F=\text{Im}\left[\frac{(1+\chi_{iL})\chi_{eL}}{1+\chi_{iL}+\chi_{eL}}\right]\cdot\frac{k_L^2 v_e^2}{\omega_{pe}^2}
\end{equation}
and the electron-cyclotron term $b_e$ is neglected since $b_e\ll 1$. The toroidal refraction rate of the pump LH wave $n_{0z}$ does not appear in Eq.\eqref{scaleF} because the experimental setting of $n_{0z}$ among different tokamaks usually satisfy $n_c+0.5<n_{0z}<n_c+1$ where $n_c=\sqrt{1+\omega_{pe}^2/\omega_{ce}^2}$ is the critical refraction rate for the accessibility of LH waves into core plasmas. The group velocity $v_{0gx}$ (and consequently, $|\phi_0|^2$) is insensitive to $n_{0z}$ at such region, thus the variation of $n_{0z}$ has minimal impact on the PDI process.

\begin{figure}[h]
    \centering
    \includegraphics[width=0.8\linewidth]{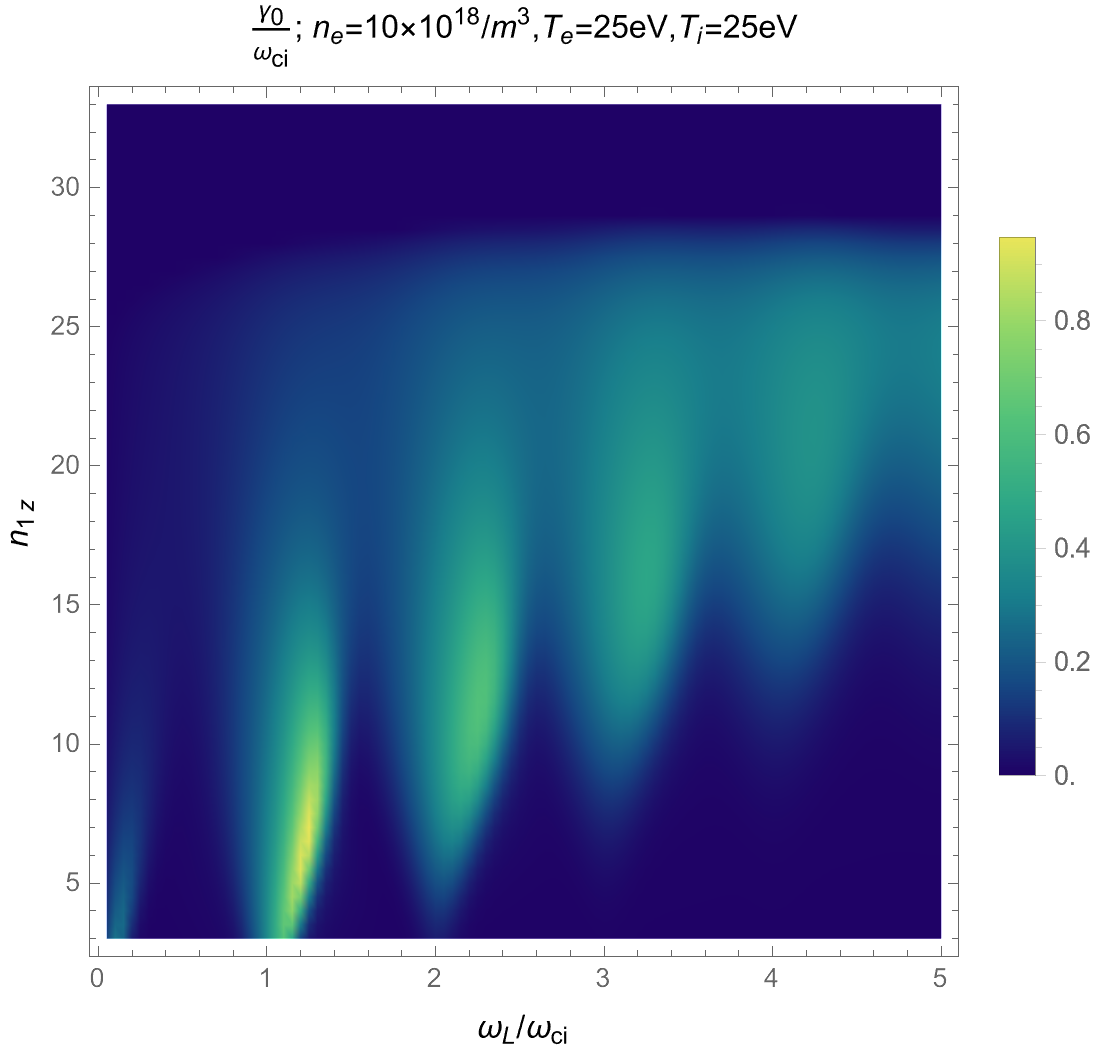}
    \caption{PDI decay channels at typical JET SOL plasma parameters ($n_e=10^{19}\text{m}^3$, $T_i=T_e=25\text{eV}$, $B_0=2.6\text{T}$) and LH antenna settings ($P_0=2\text{MW}$, $f_0=3.7\text{GHz}$, $n_{0z}=1.84$, $L_y=0.348\text{m}$, $L_z=0.884\text{m}$)}
    \label{fig3}
\end{figure}

Numerical techniques are needed to evaluate the effects of plasma parameters ($n_e,T_e,B_0$) and LH pump wave parameters ($\omega_0$) on $A_\text{amp}$, for the decay channel $n_{1z}$ as well as the value of function $F$ could be sensitive to those parameters. Thus we turn to numerical analysis based on the SOL and LH parameters of JET\cite{Cesario_2006} to complete the scaling law of $A_\text{amp}$, and consequently $n_\text{lim}$. FIG.~\ref{fig3} shows the spectrum of $\gamma_0$ using JET parameters, from which we find the channel of $\omega_L\approx\omega_{ci}$ has the highest linear growth rate. Therefore, we choose this channel to evaluate $n_\text{lim}$ in the following calculations.

The refraction factor $n_{1z}$ in Eq.\eqref{scaleF} can be treated as a function of $\zeta_s^{(n)}$ and $b_i$. Noticing that $\zeta_s^{(n)}\propto \frac{B_0}{\omega_0 n_{1z} \sqrt{T_e}}$ and $b_i \propto \frac{\omega_{LH}^2 n_{1z}^2 T_e}{B_0^2}$, we find $n_{1z}$ and $F$ share the same independent variables $\left(\omega_{LH},\omega_0,\frac{n_{1z}\sqrt{T_e}}{B_0}\right)$. The effects of the independent variables on both functions are evaluated numerically, thus we find
\begin{equation}
    n_{1z}\propto \frac{B_0}{\sqrt{T_e}}, \quad F\propto \omega_0^{-1}
    \label{n1zscale}
\end{equation}

Applying Eq.\eqref{n1zscale} to Eq.\eqref{scaleF} and considering $\omega_{LH}\propto \sqrt{n_e}$ for typical SOL parameters, we obtain the scaling laws of amplification factor
\begin{equation}
    A_\text{amp}\propto \omega_{LH}^3 P_0 L_y^{-1} T_e^{-3/2} \omega_0^{-3} B_0^{-1}
    \label{scaleA}
\end{equation}
and the density limit
\begin{equation}    
    n_\text{lim}\propto L_y^{2/3} P_0^{-2/3}\omega_0^{2}B_0^{2/3}T_e
    \label{scalen}
\end{equation}

Similar experimental results as the scaling laws Eq.\eqref{scaleA} and Eq.\eqref{scalen} have been observed in multiple tokamaks. In experiments on EAST\cite{Li_2022} and Alcator C-Mod\cite{Baek_2015}, the signals of sideband LH waves grow exponentially with plasma density. The density limits observed at early LHCD as well as electron heating experiments among different tokamaks approximately scale as $\omega_0^2$\cite{Alladio_1984}. The promotion of density limit by increasing toroidal magnetic field\cite{Li_2022,Wallace_2010} or SOL temperature\cite{Cesario_2010,Ding_2017} were also observed. Such experiments qualitatively validated the scaling laws. Furthermore, in order to provide a quantitative and more rigorous validation, we turn to simulate the process.

A ray tracing code involving LHCD induced PDI process in the SOL plasma is developed to simulate the self-consistent evolution of the spectrum and power flux of the LH wave in SOL plasma\cite{Su_phd}. Assumptions used in above analysis are freed in the simulation. Full SOL density and temperature profile is considered and the effects of both ISQM decay and ICQM decay are included. The power transfer from the pump LH wave to the sideband LH wave is calculated as Eq.\eqref{intA}, where the pump region and the sideband trajectory are determined by ray tracing. All the scatter angles are considered by sampling the sideband wave in the full wavevector space. For each step, the electric field and power flux of the pump and sideband waves are re-calculated, thus the depletion of the pump power flux is also considered along the course, and the changes of pump power and spectrum can be evaluated continuously.

\begin{figure}[h]
    \centering
    \subfigure[Evolution of LH spectrum before density limit$(n_\alpha=1.10)$]{\includegraphics[width=0.8\linewidth]{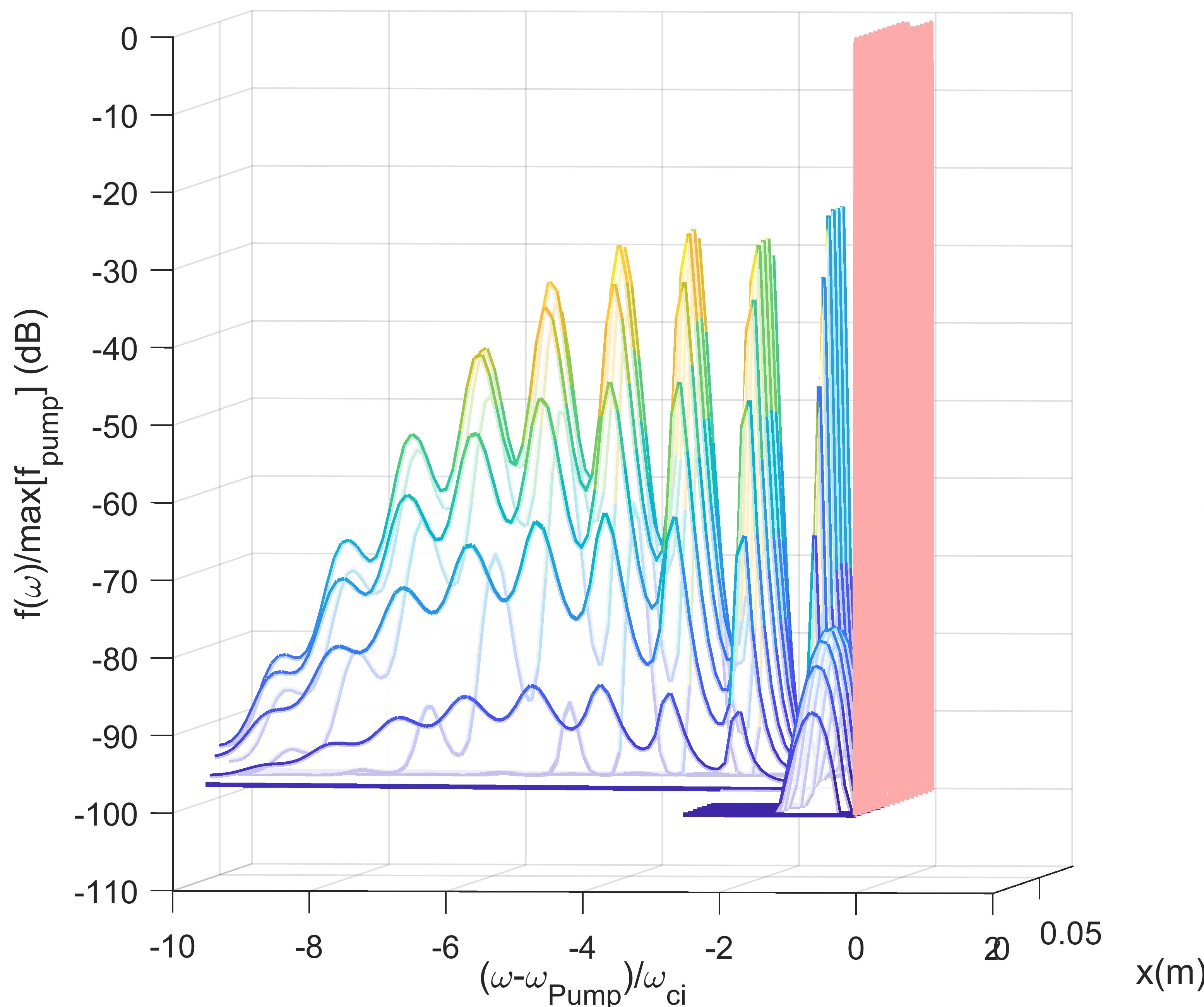}}
    \subfigure[Evolution of LH spectrum before density limit$(n_\alpha=1.19)$]{\includegraphics[width=0.8\linewidth]{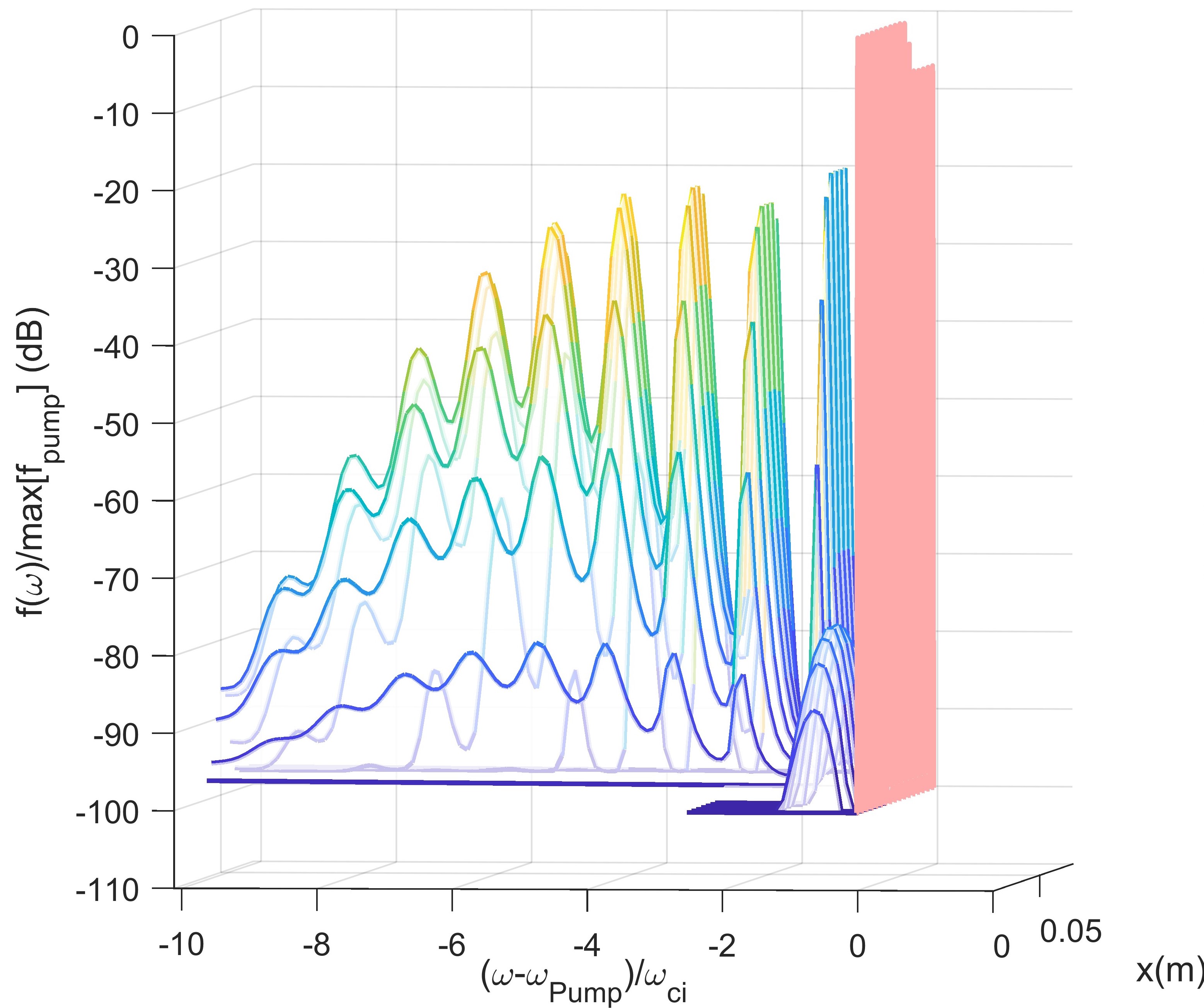}}
    \subfigure[Evolution of pump LH energy at different density\label{fig4c}]{\includegraphics[width=0.8\linewidth]{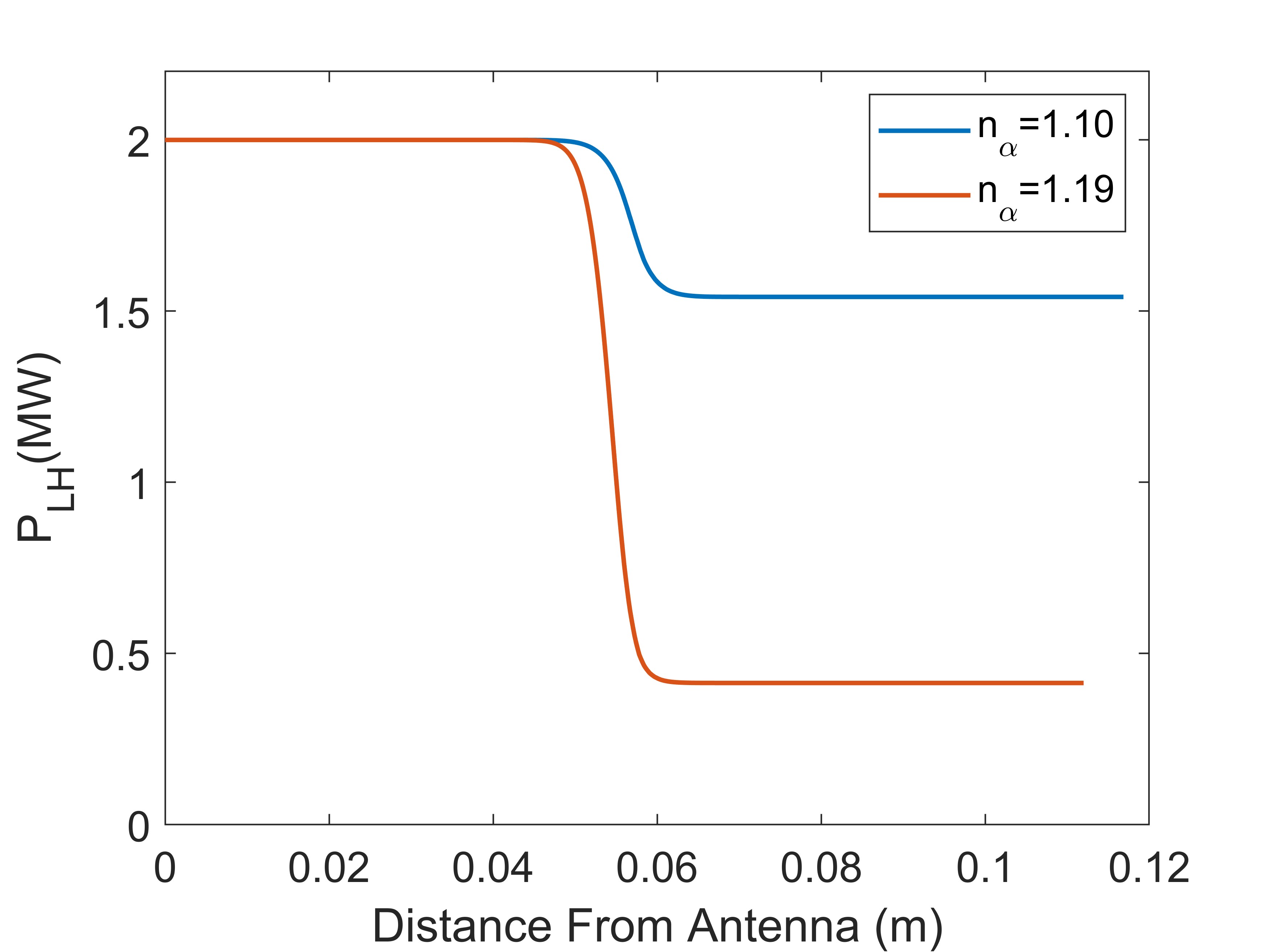}}
    \caption{Simulation results of the evolution of LH waves in SOL plasma, $x=0$ refers to the antenna mouth}
    \label{fig4}
\end{figure}

FIG.~\ref{fig4} presents two typical simulated evolution of LH spectrum and the pump power flux with strong and weak PDI convective loss. Both simulations are based on the SOL profile of JET\cite{Cesario_2006}, and the only treatment we applied is scanning the $n_e$ profile by multiplying the initial experimental profile by a factor $n_\alpha$. As shown in FIG.~\ref{fig4c}, since $n_e$ has an exponential effect on the convective loss, the density limit is reached abruptly when we modified $n_e$ by several percents ($n_\alpha$ from $1.10$ to $1.19$). Additionally, the simulation results agrees to the picture of PDI presented in above theoretical analysis. The ICQM channels dominant PDI convective loss near the density limit, and among all the ICQM decay channels, the channel with $\omega_L \approx \omega_{ci}$ has the largest amplification factor, whereas other decay channels with $2\omega_{ci}\leq\omega_L\leq 5\omega_{ci}$ are of secondary importance.

\begin{figure}[h]
    \centering
    \includegraphics[width=0.8\linewidth]{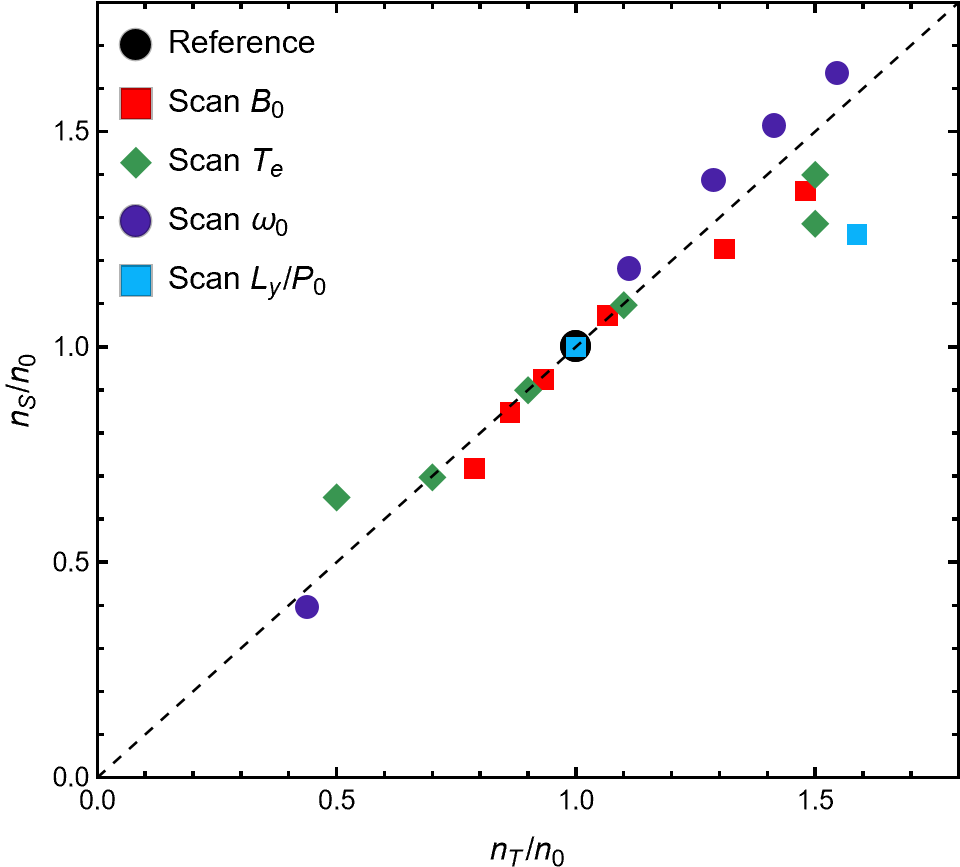}
    \caption{Comparison between simulated density limit and theoretical scaling law. Here $n_0$ is the simulated density limit at reference JET SOL profile\cite{Cesario_2006}, $n_T$ is the density limit predicted by Eq.\eqref{scaleA}, $n_S$ is the density limit obtained through simulation}
    \label{fig5}
\end{figure}

The density limit of LHCD considering the full PDI process in the SOL plasma can be evaluated through simulation by changing the factor $n_\alpha$ to modify the SOL $n_e$ profile until the PDI power loss reaches $P_\text{loss}=0.8P_0$. The simulations are performed at different plasma and LH wave parameters around the original parameters of JET (pump LH frequency $f_0$ from $2.45\text{GHz}$ to $4.6\text{GHz}$, $B_0$ from $1.8\text{T}$ to $4.0\text{T}$, average $T_e$ from $15\text{eV}$ to $30\text{eV}$), further modifications are not introduced, for they might result in effects on the position of the cut-off layer and the accessibility of the pump LH wave, thus the SOL profile of JET becomes unavailable.
The theoretical scaling law Eq.\eqref{scaleA} can be justified by comparing itself to the density limit from the simulations, as shown in FIG.~\ref{fig5}, from which we find that the theoretical and numerical results are of good agreement. The effect of modifying the antenna is slightly weaker than the prediction of the scaling law for the convective loss caused by small scatter angles $\delta<\frac{\pi}{3}$ might be stronger than that of $\delta=\frac{\pi}{2}$ at small $L_z$.


The scaling laws provides us an insight on the effectiveness of LHCD for ITER and future tokamak reactors. The expected SOL profile at the midplane of ITER 15MA baseline satisfies $n_e \sim 5\times 10^{19}/\text{m}^3$ near the LCFS\cite{Carli_2018}, thus the required density limit must be $4\sim5$ times larger compared to JET. According to the scaling law Eq.\eqref{scaleA}, the density limit of ITER is 9.55 times higher than that of JET due to higher $T_e,\omega_0$ and $B_0$, indicating that the convective loss of LH wave is minimal. Such results agrees with the simulation based on the expected ITER SOL\cite{Carli_2018} profile and previous ITER LH antenna design\cite{Belo_2015}, in which the convective loss at the SOL region is less than 1\% of the total LH wave power flux. It also indicated that the achievements of the effective LHCD reached at ITER-level high density on Alcator C-Mod\cite{Baek_2018} and FTU\cite{Cesario_2010} are solid in physics.

In conclusion, by introducing the PDI process induced by the LH wave in the SOL plasma and coupling the process to the propagation of LH waves, an effective physical model of the density limit of LHCD was established. Theoretical scaling law of the density limit in good agreement with simulation results and previous experiments among different tokamaks was acquired. Both the theoretical scaling law and the simulation results indicate that the convective loss at the SOL plasma can be minimal for ITER baseline scenarios. Several potential methods to suppress the convective loss and consequently improve the effectiveness of LHCD in high density plasma were provided. In total, the density limit will not prevent LHCD from being an efficient and promising method of driving plasma current for future fusion reactors.

\begin{acknowledgments}
This work was supported by NSFC under Grant No.12335014.
\end{acknowledgments}

\end{document}